\documentclass[prl,preprint]{revtex4-1}
\usepackage{graphicx}
\usepackage{dcolumn}
\usepackage{bm}
\usepackage{color}
\usepackage{tabularx}
\usepackage{array}
\usepackage{amsmath}

\begin{document}

\title{Inertial terms to magnetization dynamics in ferromagnetic thin films}

\author{Y. Li$^{1,4}$, A.-L. Barra$^{2}$, S. Auffret$^{3,4,5}$, U. Ebels$^{3,4,5}$, and W. E. Bailey$^{1*}$ \\
\textnormal{$^{1}$ Materials Science \& Engineering, Dept. of Applied Physics \& Applied Mathematics, Columbia University, New York NY 10027, USA\\
$^{2}$ Laboratoire National des Champs Magn\'{e}tiques Intenses, CNRS/UJF/UPS/INSA, 38042 Grenoble Cedex, France\\
$^3$ Univ. Grenoble Alpes, INAC-SPINTEC, F-38000 Grenoble, France\\
$^4$ CNRS, SPINTEC, F-38000 Grenoble, France\\
$^5$ CEA, INAC-SPINTEC, F-38000 Grenoble, France}\\
$^{*}$ Correspondence to: web54@columbia.edu
}

\date{\today}

\begin{abstract}

Inertial magnetization dynamics have been predicted at ultrahigh speeds, or frequencies approaching the energy relaxation scale of electrons, in ferromagnetic metals. Here we identify inertial terms to magnetization dynamics in thin Ni$_{79}$Fe$_{21}$ and Co films near room temperature. Effective magnetic fields measured in high-frequency ferromagnetic resonance (115-345 GHz) show an additional stiffening term which is quadratic in frequency and $\sim$ 80 mT at the high frequency limit of our experiment. Our results extend understanding of magnetization dynamics at sub-picosecond time scales.

\end{abstract}

\maketitle

\indent The magnetization $M(t)$ in ferromagnetic materials is generally understood to evolve without memory of its prior motion. As described by the Landau-Lifshitz (LL) equation\cite{llg}, magnetization dynamics $dM(t)/dt$ can be written in terms of the magnetization $M(t)$ alone, excluding any temporal derivatives $d^nM(t)/dt^n$. Magnetization is then 'inertialess': it responds to a step magnetic field $H(t)$ with an instantaneous change in speed and with infinite acceleration.\\
\indent As pointed out by Ciornai \textit{et al.} and F\"{a}hnle \textit{et al.}\cite{ciorneiPRB2011,fahnlePRB2011,fahnlePRB2011err}, followed by other theoreticians\cite{bottcherPRB2012,bhattacharjeePRL2012}, the absence of inertia is questionable for magnetization dynamics at very high frequencies. The high frequency behavior, $>$ 100 GHz, becomes relevant in ultrafast ($<$ ps-scale) magnetization switching schemes envisaged for magnetic information storage technology\cite{backScience1999,tudosaNature2004,stanciuPRL2007,kimelnphys2009, manginNMAT2014}. In metallic ferromagnets, the electrons which comprise the magnetization themselves possess inertia, expressed through a finite (Bloch-state) lifetime $\tau$ and cannot change their momenta infinitely quickly. These authors\cite{ciorneiPRB2011,fahnlePRB2011,fahnlePRB2011err} have proposed that the LL equation should be amended to include an inertial term:
\begin{equation}
{d\mathbf{m} \over dt}=-\mu_0\gamma [\mathbf{m}\times\mathbf{H_{eff}} + \alpha \mathbf{m}\times(\mathbf{m}\times\mathbf{H_{eff}})]
\pm \alpha \mathbf{m}\times \tau{d^2\mathbf{m} \over dt^2}
\end{equation}
Here $\mathbf{m}$ is the unit magnetization vector, $\gamma$ is the gyromagnetic ratio, $\mathbf{H_{eff}}$ is the effective field, and $\alpha$ is the Gilbert damping coefficient. The additional nonlinear term $\pm \alpha \mathbf{m}\times \tau d^2\mathbf{m} / d^2t$ is second-order in time. It is straightforward to show that this inertial term leads to an effective field $H_{eff}' = \mp \alpha\omega^2\tau/\gamma$, quadratic in frequency, which acts on magnetization dynamics. Increasing the frequency limit for ferromagnetic resonance (FMR) experiments by a factor of 10 will therefore increase the magnitude of the inertial effective field by a factor of 100. \\
\indent The sign of the inertial term in Eq. (1) refers to two different origins for the proposed inertial dynamics. Ciornai \textit{et al.}\cite{ciorneiPRB2011,wegroweAJP2012} argue that additional energy terms for the precession, typically neglected, come about from transverse magnetization motion. Finite rotational moments of inertia about the two transverse axes were introduced semiclassically to account for this energy in Ref. \cite{ciorneiPRB2011}. On the other hand, F\"{a}hnle \textit{et al.}\cite{fahnlePRB2011,fahnlePRB2011err} have pointed out that similar dynamics arise from a more complete consideration of damping. Here the inertia can be associated with the linear momentum of electrons: electrons cannot change their momenta $\bf{k}$ more rapidly during precession than their Bloch state relaxation time $\tau_B$, and tend to remain at higher energy states. In this sense, inertia is proportional to $\tau_B$. The {\it rotational} moment of inertia ("+" in Eq. 1) softens the resonance frequency while the {\it linear} inertia due to relaxation ("-" in Eq. 1) stiffens the resonance frequency. \\
\indent In this work, we test for the existence of inertial effective fields which act on the magnetization of the device ferromagnets permalloy (Ni$_{79}$Fe$_{21}$, or Py) and Co. To do so, we measure FMR up to 345 GHz, roughly an order of magnitude higher than is accessible using conventional sources. Through measurements of the frequency-dependent resonance field $\mu_0H_{res}(\omega)$, we find in all samples a negative nonlinear shift quadratic in frequency which reflects a positive effective field added by the inertial term. The sign of the shift shows the dominance of inertia due to the finite Bloch state relaxation time\cite{fahnlePRB2011,fahnlePRB2011err} over the inertia due to finite rotational moment of inertia of electrons\cite{ciorneiPRB2011}. In all samples the extracted values of $\tau$ are comparable with low-fluence limits of remagnetization times $\tau_E$ measured in ultrafast demagnetization experiments. Our results show that high-frequency FMR (HF-FMR) provides a near-ground-state ($\sim$ 1 meV) view of the processes involved in ultrafast magnetization dynamics, heretofore accessible through far-from-equilibrium techniques.\\
\begin{center}
\begin{figure*}[htb]
\centering
\includegraphics[width = 6.0 in] {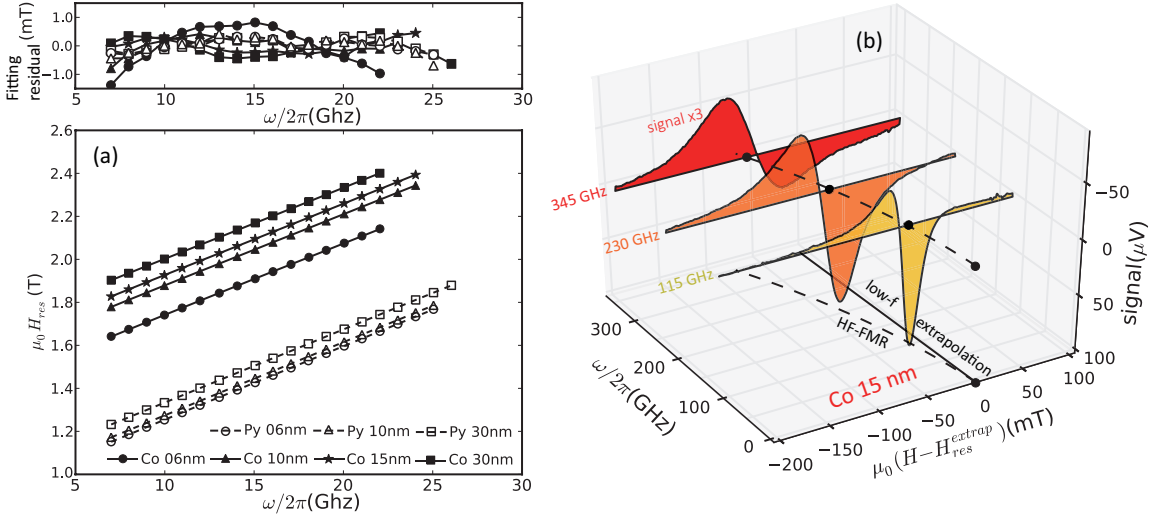}
\caption{ (a) Low-frequency resonance fields for Py and Co $<$ 26 GHz. Lines and dashed lines are fits to ${\omega/\gamma}=\mu_0(H_{res}-M_{eff})$. The upper panel shows the fitting residuals of the lower panel. (b) High-frequency FMR absorption spectra for Co 15 nm. The (horizontal) field axis shows the difference of measured high-frequency resonance fields with expected values based on extrapolation of the low-frequency resonance fields without inertial effects. The solid data points indicate the resonance field $\mu_0H_{res}$, with fits shown in the dashed curve.}
\end{figure*}
\end{center}
\indent Py and Co thin films were deposited on Si/SiO$_2$ substrates using ultra-high vacuum magnetron sputtering. The capping layers are Cu(5 nm)/SiO$_2$(5 nm) for Py and Cu(5 nm)/Ta(3 nm) for Co; the seed layers are Ta(5 nm)/Cu(5 nm) in both series. The thicknesses $t_{FM}=$ 6, 10 and 30 nm for Py and 6, 10, 15, 30 nm for Co, respectively. Two series of field-swept, variable-frequency FMR experiments were carried out, both with applied field and static magnetization perpendicular to film plane. At low fields and frequencies (3-26 GHz, 1-2.5 T, 298 K), a coplanar waveguide was used to measure the power transmission through the thin films. These measurements were carried out at Columbia University using an Fe-core electromagnet\cite{yiarXiv2013}. At high fields and frequencies (115, 230 and 345 GHz, 4-13 T, 273-280 K) continuous-wave high-field electron paramagnetic resonance was used to measure the power reflection from the samples. These measurements were carried out at the LNCMI-G using a superconducting magnet\cite{barraAMR2006}. Six field sweeps were taken for HF-FMR to reduce measurement and hysteresis errors.\\
\indent Several steps have been taken to ensure the consistency of low- and high-frequency measurements. These are essential to detect the nonlinear term in magnetization dynamics. First, the magnetic fields of both low- and high-frequency FMR were calibrated \textit{in-situ} using paramagnetic markers\cite{fieldcali}. An accuracy of 0.01\% is obtained, corresponding to $\sim$ 1 mT at 345 GHz ($\sim$ 13 T). Second, for low-frequency FMR, each sample normal has been independently aligned along the applied field by rotating the waveguide with respect to two axes (with $<0.2^\circ$ precision) to maximize the value of $\mu_0H_{res}$ at 3 GHz. The samples were aligned optically, to lower precision, in HF-FMR. Third, we have corrected for the small ($\sim$ 20 K) temperature differences between the two apparati, through variable-temperature FMR measurements at Columbia University, which allow us to correct for temperature-dependent magnetization $\mu_0M_{eff}(T)$. See the Supplemental Information for details.\\
\indent Fig. 1(a) shows the low-frequency $\mu_0H_{res}$ as a function of $\omega/2\pi$ for Py and Co. Continuous and dashed lines are fits to the linear Kittel equation ${\omega/\gamma}=\mu_0(H_{res}-M_{eff})$. The top panel shows the fitting residuals, which are randomly distributed within $\pm$1 mT. In order to examine whether resonance field offset exists at high frequencies, we plot in Fig. 1(b) the HF-FMR lineshapes (average of six measurements) for Co 15 nm, as a function of $\mu_0(H-H_{res}^{extrap})$, with $\mu_0H_{res}^{extrap}$ the resonance field extrapolated from low-frequency FMR. The solid circles indicate the resonance field $\mu_0H_{res}$ extracted from the FMR lineshapes for the three frequencies. The dashed lines are a fit of the resonance positions to a quadratic nonlinear term to frequency. A negative curvature is found in the dashed lines. This implies a negative sign in Eq. (1) with positive effective inertial field. Similar effects are found in other Py and Co films. \\
\begin{center}
\begin{figure*}[htb]
\centering
\includegraphics[width = 6.0 in] {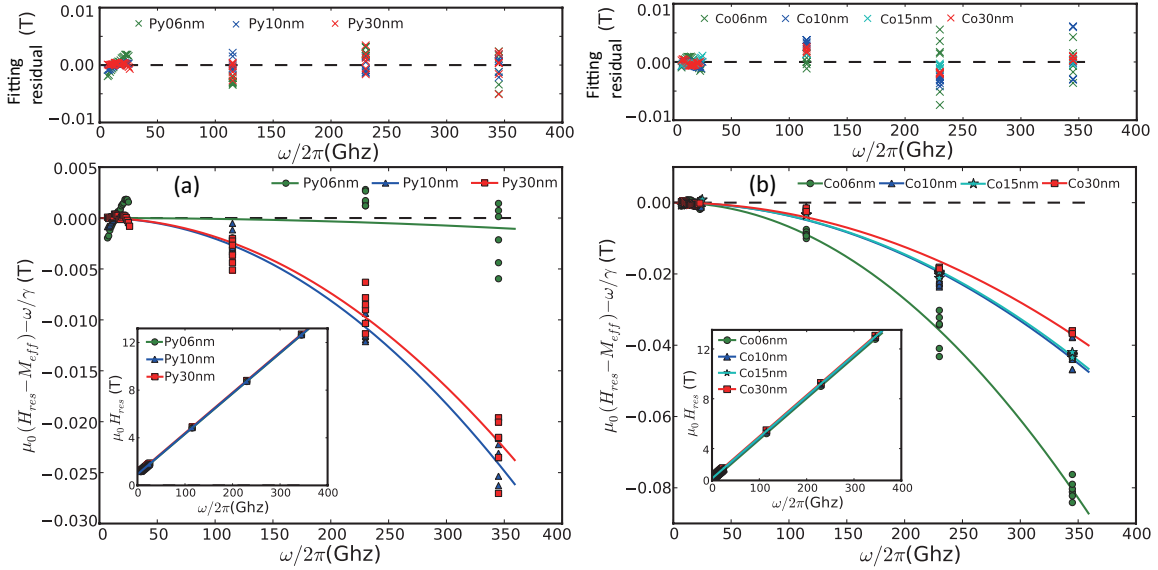}
\caption{Deviation of $\mu_0H_{res}$ from the linear Kittel equation for (a) Py and (b) Co. Main panels: $\mu_0H_{res}|_{exp}-(\mu_0M_{eff} + \omega/\gamma)|_{fit}$ in solid data points; $\alpha_0\omega^2\tau/\gamma|_{fit}$ in curved lines. $M_{eff}$, $\gamma$ and $\tau$ are the fit parameters from Eq. (2). Upper panels: fitting residuals of main panels as a function of frequency. \textit{Insets}: full plots of $\mu_0H_{res}$ and fits to Eq. (2).}
\end{figure*}
\end{center}
\indent To quantify the nonlinear term from Eq. (1), we fit the resonance fields of both low- and high-frequency FMR measurements together into the linear Kittel equation, plus a quadratic term:
\begin{equation}
\mu_0H_{res} = \mu_0M_{eff} + \omega/\gamma - \alpha_0\omega^2\tau/\gamma
\end{equation}
taking $M_{eff}$, $\gamma$ and $\tau$ as fit parameters. The parameter $\tau$ governs the magnitude of the inertial term. The Gilbert damping coefficients $\alpha_0$ are extracted by fitting the low-frequency FMR linewidths into $\mu_0\Delta H_{1/2} = \mu_0\Delta H_0 + 2\alpha_0\omega/\gamma$, where $\mu_0\Delta H_0$ is the inhomogeneous linewidth. To estimate errorbars for HF-FMR, we take the scatter of all six measurements into the data sequence. \\
\indent Figure 2 contains the central result of this manuscript. In the main panels we show experimental resonance fields $\mu_0H_{res}$ and fits to Eq. (2) as a function of frequency $\omega/2\pi$ for (a) Py and (b) Co. The conventional effective field terms $\mu_0 M_{eff} + \omega/\gamma$ have been subtracted away; plotting the data as $\mu_0H_{res}|_{exp}-(\mu_0M_{eff} + \omega/\gamma)|_{fit}$ (solid data points) and $\alpha_0\omega^2\tau/\gamma|_{fit}$ (curves) isolates the inertial contribution. The fit values are tabulated in Table 1; error bars denote $3\sigma$ of uncertainty. We find a reasonable fit to the data through the introduction of the quadratic term. For both Py and Co, the values of $\tau$ stay roughly at the same level between 10-30 nm: $\sim 0.1$ ps for Py and $\sim 0.3$ ps for Co. For Py 06 nm negligible shifts in $\mu_0H_{res}$ is found, as will be discussed. \\
\begin{center}
\begin{table*}[ht]
\centering
\begin{tabular}{>{\centering\arraybackslash}m{0.8in} | >{\centering\arraybackslash}m{0.8in} >{\centering\arraybackslash}m{0.8in} >{\centering\arraybackslash}m{0.8in} >{\centering\arraybackslash}m{1.0in} >{\centering\arraybackslash}m{1.0in} }
Sample & $\mu_0M_{eff}$ & $g_{eff}$ & $\alpha_0$ & $\tau$($\mu_0H_{res}$) & $\tau$($\mu_0\Delta H_{1/2}$) \\
\hline
\hline
Py 06 nm & 0.915 T & 2.106 & 0.0075 & $^*$0.28$\pm$0.06 ps & 0.09$\pm$0.09 ps\\
Py 10 nm & 0.935 T & 2.100 & 0.0072 & 0.13$\pm$0.03 ps & 0.11$\pm$0.04 ps\\
Py 30 nm & 0.993 T & 2.101 & 0.0069 & 0.12$\pm$0.05 ps & 0.06$\pm$0.08 ps\\
\hline
\hline
Co 06 nm & 1.409 T & 2.147 & 0.0067 & 0.47$\pm$0.08 ps & 0.19$\pm$0.07 ps\\
Co 10 nm & 1.547 T & 2.153 & 0.0053 & 0.26$\pm$0.09 ps & 0.14$\pm$0.07 ps\\
Co 15 nm & 1.596 T & 2.154 & 0.0051 & 0.32$\pm$0.03 ps & 0.15$\pm$0.05 ps\\
Co 30 nm & 1.672 T & 2.155 & 0.0056 & 0.26$\pm$0.05 ps & 0.21$\pm$0.09 ps\\
\hline
\end{tabular}
\caption{Fit parameters of experimental resonance fields $\mu_0H_{res}(\omega)$ to the inertial term Eq. (2) and linewidths $\mu_0\Delta H_{1/2}$ to Eq. (3) for Py and Co. Values of low-frequency Gilbert damping $\alpha_0$ are found from linear fits of low-frequency linewidths. Errorbars of $\tau$ represent $3\sigma$ deviations. $^*$Misaligment analysis has been applied; see text for details.}
\label{table1}
\end{table*}
\end{center}
\indent A primary source of uncertainty may come from sample misalignment in HF-FMR measurements. Given the experimental geometry, the samples cannot be aligned independently of mounting. Canting the sample off normal alignment in HF-FMR will also lead to reduced resonance fields as in Fig. 1(b). Misalignment will be manifested as a \textit{linear} deviation of $\mu_0H_{res}|_{exp}-(\mu_0M_{eff} + \omega/\gamma)|_{fit}$ from being zero at low frequencies ($\omega/2\pi<30$ GHz) in Fig. 2. For Co samples, the deviations at low frequencies are negligible, indicating that Co samples are well aligned perpendicular to biasing field. For Py the same is true, with the exception of the 6 nm sample. This sample shows a significant linear deviation at low frequencies (green circles), indicating that the sample plane may not be perfectly perpendicular to the biasing field. According to simulations, shown in the Supplemental Information, such a deviation would be produced by a misalignment of $\sim 6^\circ$ for Py 06 nm, which will add an additional negative $\mu_0H_{res}$ shift of $\sim20$ mT. With this correction, the estimated $\tau$ is increased from 0.01 ps to 0.28 ps in Py 06 nm, consistent with the larger value for Co 6 nm. The corrections are much smaller for other samples and have been neglected.\\
\indent The influence of the inertial dynamics which we observe is a stiffening, rather than a softening, of the resonance. We conclude therefore that the inertial term from {\it linear} momentum\cite{fahnlePRB2011,fahnlePRB2011err,bhattacharjeePRL2012} dominates over the inertial term from {\it angular} momentum proposed in Ref. \cite{ciorneiPRB2011}. The former term is a direct product of the breathing Fermi surface model causing conductivity-like Gilbert damping\cite{kamberskyCJP1970,kamberskyPRB2007}. In the conductivity-like model the value of observed $\tau$, 0.1-0.5 ps, represent the average of Bloch state lifetime $\tau_B$ near the Fermi surface for Py and Co.\\
\begin{center}
\begin{figure*}[htb]
\centering
\includegraphics[width = 6.0 in] {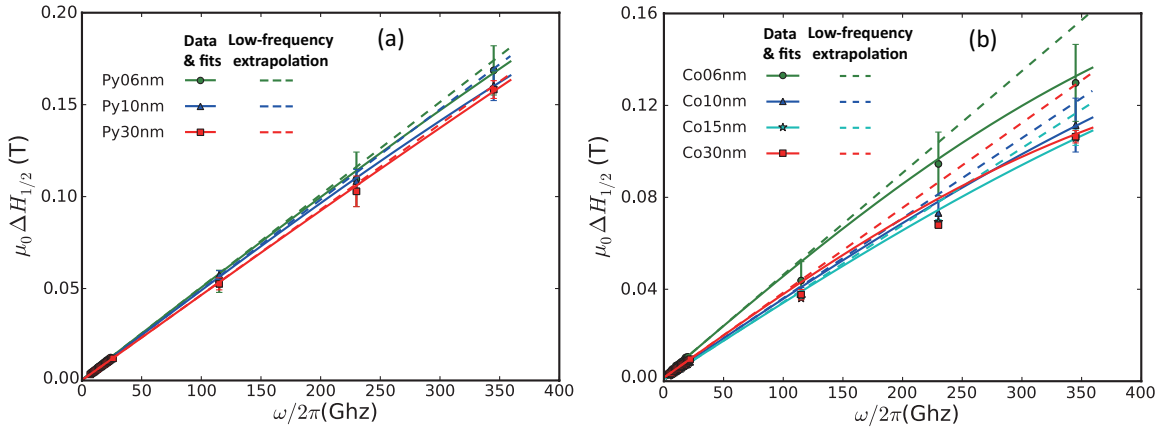}
\caption{Ferromagnetic resonance half-power linewidths $\mu_0\Delta H_{1/2}$ as a function of $\omega/2\pi$ for (a) Py and (b) Co, between 4 to 345 GHz. Dashed lines are linear extrapolations of the low-frequency FMR linewidths; solid curves are fits to Eq. (3). Each point with error bar represents a summary of six measurements.}
\end{figure*}
\end{center}
\indent Note that Gilbert damping coefficient $\alpha$, calculated by the breathing Fermi surface model, should also be frequency-dependent. A relation of $\alpha = \alpha_0/(1+\omega^2\tau_B^2)$ can be obtained\cite{kamberskyCJP1970,kamberskyPRB2007,clogstonBSTJ1955} with $\alpha_0$ the zero-frequency damping. Fig. 3 shows the linewidths of both low-frequency and high-frequency ranges for Py and Co. Here points with errorbars (one $\sigma$) are used for frequencies $>$ 100 GHz to reflect the scattering of six repeats. The dashed lines are low-frequency linear extrapolations from $\alpha_0$. Compared with extrapolated linewidths from the low-frequency $\Delta H_{1/2}(\omega)$, we observe reduced linewidths at high frequencies (115-345 GHz). No explicit prediction has been made for the effect of rotational inertia on the linewidth in Ref. \cite{ciorneiPRB2011,wegroweAJP2012}, but the observed behavior matches well with the prediction of linear inertial terms in Ref. \cite{kamberskyCJP1970,kamberskyPRB2007,clogstonBSTJ1955}. The solid curves are fits to the form:
\begin{equation}
\mu_0\Delta H_{1/2} = \mu_0\Delta H_0 + {2\alpha_0\omega\over\gamma}{1\over 1+\omega^2\tau^2}
\end{equation}
taking the $\alpha_0$ from the low-frequency linewidths. The fitted $\tau$ are listed in Table 1. The values of $\tau$ extracted from $\mu_0\Delta H_{1/2}$ are close but slightly smaller than from $\mu_0H_{res}$. We note here that the linewidth measurements are less precise than $\mu_0H_{res}$ and more sensitive than resonance fields to various sources of inhomogeneities in the samples. \\
\indent At room temperature, the relaxation rate of Bloch states ($1/\tau_B$) is determined by the electron scattering with phonons and impurities. In this sense, $\tau_B$ is similar to the remagnetization time $\tau_E$ in the ultrafast demagnetization experiments\cite{beaurepairePRL1996,koopmansPRL2005,koopmansNMAT2009}, where optically excited electrons also relax through electron-phonon and electron-impurity interactions. In the limit of zero laser fluence, nonlinear effects due to high occupation number of excited states are reduced\cite{boscoAPL2003,koopmansNMAT2009}. The zero-fluence $\tau_E$ has been reported to be 0.2-0.25 ps for Py 10-30 nm\cite{boscoAPL2003} and $\sim 0.4$ ps for Co 15 nm\cite{koopmansNMAT2009}, close to the value of $\tau$ in this work. \\
\indent At high frequencies the (nonmagnetic) skin depths $\delta_s$ in the ferromagnetic films become smaller and the enhanced eddy current effect may influence the resonance field\cite{ppincus1960}. The resonance field will be enhanced by $\sim\mu_0M_s / \delta_s^4k_0^4$, where $k_0$ is the lowest-energy wavenumber determined by the surface anisotropy. Because $\delta_s^2\propto1/\omega$, the resonance field enhancement is proportional to $\omega^2$ and may influence the quadratic term in Eq. (2). Our calculations show that this term is negligible, about 0.4 mT for Py 30 nm and 0.09 mT for Co 30 nm, compared with the observed effects up to 80 mT (See the Supplemental Information for details).\\
\indent We do not believe that interfacial effects, including spin pumping, play an important role in the observed high-frequency behavior. Both Gilbert damping $\alpha_0$ and the two inertial dynamics lifetimes $\tau (\mu_0H_{res})$ and $\tau (\mu_0\Delta H_{1/2})$ in Table 1 show little thickness dependence for either Py or Co, indicating that bulk relaxation is dominant. The weak thickness dependence of $\alpha_0$ is consistent with the very low spin pumping effect of Py/Ta identified in Ref. \cite{mizukamiJMMM2001}, in any case negligible for 30 nm films and without quadratic frequency dependence. Theoretical predictions of resonance shifts from imaginary spin mixing conductance are three orders of magnitude lower than observed here\cite{zwierzyckiPRB2005}. Only $\tau (\mu_0 H_{res})$ measured from resonance shifts for 6 nm films, not matched in $\tau (\mu_0 \Delta H_{1/2})$ measured from linewidths, differ significantly. This enhancement may be structural in origin.\\
\indent A technological implication of our results is that effective field requirements for precessional switching will be \textit{reduced} as switching times in magnetic storage decrease into the few-picosecond range. In this sense, the inertial dynamics ease ultrafast switching, if the behaviors of Py and Co are representative of other metallic ferromagnets. The effective field reduction, here up to 80 mT in Co, is not small in an absolute sense, and might according to Eq. (2) be enhanced significantly in ferromagnets with higher Gilbert damping. On the other hand, prior switching experiments with high-field relativistic electron bunches seem to indicate that nonlinear damping increases effective field requirements by a rather larger amount for large-angle dynamics in CoCrPt\cite{tudosaNature2004}, underscoring the utility of HF-FMR to identify the inertial effect.\\
\indent In summary, we identify a novel term to magnetization dynamics in the ferromagnetic metal films Py and Co which is quadratic in frequency and becomes significant above 100 GHz. The term stiffens the frequency, aiding applied fields in driving ultrafast magnetization motion. The behavior is best explained by dynamics retarded through a finite Bloch-state relaxation time $\tau_B$ as proposed in Refs. \cite{fahnlePRB2011,fahnlePRB2011err,bhattacharjeePRL2012}. Extracted relaxation times are 0.1-0.2 ps for Py and 0.2-0.4 ps for Co, close to the remagnetization times measured in optical pump-probe demagnetization experiments. Our findings extend understanding of magnetization dynamics at picosecond time scales and may open up new possibilities for high-speed inertial switching in ferromagnetic materials, previously demonstrated only in antiferromagnets\cite{kimelnphys2009}.\\
\indent We thank M. F\"{a}hnle, J. E. Wegrowe and J. Fransson for discussions and X. Yang for suggestions on statistical analysis. We acknowledge fundings by the EC through CRONOS (N$^\circ$ 280829), NSF-DMR-1411160, and the Chair of Excellence Program of the Nanosciences Foundation in Grenoble France for support.
\bibliographystyle{ieeetr}

\begin{thebibliography}{5}


\bibitem{llg}
Eq. (1) is equivalent to the Landau-Lifshitz-Gilbert (LLG) equation with renormalized gyromagnetic ratio $\gamma$. See:
T. L. Gilbert,
\textit{IEEE Trans. Magn.} \textbf{6}, 3443 (2004).

\bibitem{ciorneiPRB2011}
M.-C. Ciornei, J. M. Rubi and J.-E. Wegrowe,
\textit{Phys. Rev. B} \textbf{83}, 020410(R) (2011).

\bibitem{fahnlePRB2011}
M. F\"{a}hnle, D. Steiauf and C. Illg,
\textit{Phys. Rev. B} \textbf{84}, 172403 (2011).

\bibitem{fahnlePRB2011err}
M. F\"{a}hnle, D. Steiauf and C. Illg,
\textit{Phys. Rev. B} \textbf{88}, 219905(E) (2013).

\bibitem{bottcherPRB2012}
D. B\"{o}ttcher and J. Henk,
\textit{Phys. Rev. B} \textbf{86}, 020404(R) (2012).

\bibitem{bhattacharjeePRL2012}
S. Bhattacharjee, L. Nordstr\"{o}m and J. Fransson,
\textit{Phys. Rev. Lett.} \textbf{108}, 057204 (2012).

\bibitem{backScience1999}
C. H. Back, R. Allenspach, W. Weber, S. S. P. Parkin, D. Weller, E. L. Garwin and H. C. Siegmann,
\textit{Science} \textbf{285}, 864 (1999).

\bibitem{tudosaNature2004}
I. Tudosa, C. Stamm, A. B. Kashuba, F. King, H. C. Siegmann, J. St\"{o}hr, G. Ju, B. Lu and D. Weller,
\textit{Nature} \textbf{428}, 831 (2004).

\bibitem{stanciuPRL2007}
C. D. Stanciu, F. Hansteen, A. V. Kimel, A. Kirilyuk, A. Tsukamoto, A. Itoh and Th. Rasing,
\textit{Phys. Rev. Lett.} \textbf{99}, 047601 (2007).

\bibitem{kimelnphys2009}
A. V. Kimel, B. A. Ivanov, R. V. Pisarev, P. A. Usachev, A. Kirilyuk and Th. Rasing,
\textit{Nature Phys.} \textbf{5}, 727 (2009).


\bibitem{manginNMAT2014}
S. Mangin, M. Gottwald, C.-H. Lambert, D. Steil, V. Uhl\'{i}\v{r}, L. Pang,
M. Hehn, S. Alebrand, M. Cinchetti, G. Malinowski, Y. Fainman, M. Aeschlimann and E. E. Fullerton,
\textit{Nature Mater.} \textbf{13}, 286 (2014).

\bibitem{wegroweAJP2012}
J.-E Wegrowe and M.-C. Ciornei,
\textit{Am. J. Phys.} \textbf{80}, 607 (2012).

\bibitem{yiarXiv2013}
Y. Li and W. E. Bailey,
\textit{arXiv} 1401.6467

\bibitem{barraAMR2006}
A.-L. Barra, A. K. Hassan, A. Janoschka, C. L. Schmidt and V. Sch\"{u}nemann,
\textit{Appl. Magn. Reson.} \textbf{30}, 385 (2006).

\bibitem{fieldcali}
Paramagnetic markers used: BDPA complex with benzene(1:1)
free radical for low-frequency FMR, $g_{eff}$=2.0025; MnO diluted in the
diamagnetic host MgO for high-frequency FMR, $g_{eff}$=2.00101

\bibitem{kamberskyCJP1970}
V. Kambersk\'{y},
\textit{Can. J. Phys.} \textbf{48}, 2906 (1970).

\bibitem{kamberskyPRB2007}
V. Kambersk\'{y},
\textit{Phys. Rev. B} \textbf{76}, 134416 (2007).

\bibitem{clogstonBSTJ1955}
A. M. Clogston,
\textit{Bell Syst. Tech. J.} \textbf{34}, 739 (1955).

\bibitem{beaurepairePRL1996}
E. Beaurepaire, J.-C. Merle, A. Daunois and J.-Y. Bigot,
\textit{Phys. Rev. Lett.} \textbf{76}, 4250 (1996).

\bibitem{koopmansPRL2005}
B. Koopmans, J. J. M. Ruigrok, F. Dalla Longa and W. J. M. de Jonge,
\textit{Phys. Rev. Lett.} \textbf{95}, 267207 (2005).


\bibitem{koopmansNMAT2009}
B. Koopmans, G. Malinowski, F. Dalla Longa, D. Steiauf, M. F\"{a}hnle, T. Roth, M. Cinchetti and M. Aeschlimann,
\textit{Nature Mater.} \textbf{9}, 259 (2009).

\bibitem{boscoAPL2003}
C. A. C Bosco, A. Azevedo and L. H. Acioli,
\textit{Appl. Phys. Lett.} \textbf{83}, 1767 (2003).

\bibitem{ppincus1960}
P. Pincus,
\textit{Phys. Rev.} \textbf{118}, 658 (1960).

\bibitem{mizukamiJMMM2001}
S. Mizukami, Y. Ando and T. Miyazaki,
\textit{J. Magn. Magn. Mater.} \textbf{226}, 1640 (2001).

\bibitem{zwierzyckiPRB2005}
M. Zwierzycki, Y. Tserkovnyak, P. J. Kelly, A. Brataas and G. E. W. Bauer,
\textit{Phys. Rev. B} \textbf{71}, 064420 (2005).


\end{thebibliography}

\end{document}